# A FAREY FRACTION SPIN CHAIN

P. Kleban*
Department of Physics and Astronomy
and
Laboratory for Surface Science and Technology
University of Maine
Orono, ME 04469
USA

A. E. Özlük
Department of Mathematics and Statistics
University of Maine
Orono, ME 04469
USA

We introduce a new number–theoretic spin chain and explore its thermodynamics and connections with number theory. The energy of each spin configuration is defined in a translation–invariant manner in terms of the Farey fractions, and is also expressed using Pauli matrices. We prove that the free energy exists and exhibits a unique phase transition at inverse temperature $\beta = 2$. The free energy is the same as that of a related, non translation–invariant number–theoretic spin chain. Using a number–theoretic argument, the low–temperature ($\beta > 3$) state is shown to be completely magnetized for long chains. The number of states of energy $E = \log(n)$ summed over chain length is expressed in terms of a restricted divisor problem. We conjecture that its asymptotic form is (n log n), consistent with the phase transition at $\beta = 2$, and suggesting a possible connection with the Riemann $\zeta$–function. The spin interaction coefficients include all even many–body terms and are translation invariant. Computer results indicate that all the interaction coefficients, except the constant term, are ferromagnetic.

* kleban@maine.edu

## I. Introduction

There has been considerable recent interest in the area of statistical mechanical models inspired by or closely connected with number theory. An overall goal of this work is to illuminate connections between the two disciplines, so that each might provide new insights and techniques useful for the other. There is, more specifically, the hope that a direct link between the Riemann hypothesis and the Lee–Yang theory of phase transitions, based on zeros of the partition function, will eventually emerge.

In this paper, we introduce and examine a new equilibrium statistical mechanics spin chain model based on the number–theoretic Farey fractions. Our model is closely related to the number–theoretic spin chain studied by Knauf (see references given below) and others (see [Cv]), and in fact has the same free energy. However it differs from that previous work in several respects. Perhaps most important, our model is translation invariant by construction. In addition, our matrix formulation clarifies some of the results on the number theoretic spin chain. However, we have yet to conclusively demonstrate a connection with the Riemann $\zeta$–function, as is the case (by construction) for the previous model at low temperatures.

In Section II we define the new model, and point out its connection to the number theoretic spin chain studied by Knauf and others, which we refer to as KSC below. We also exhibit exact results for the partition function at certain temperatures. Section III contains a proof that the free energy (of the infinite system) exists and has a unique phase transition at $\beta = 2$. In fact it is equal to the KSC free energy, establishing a direct connection between the two models. In Section IV we examine the number of states, and show it is related to a certain number–theoretic restricted divisor problem. Using a conjectured asymptotic form for the summed (over all chain lengths) number of states we then suggest a connection with the Riemann function. Section V uses the results of the previous Section to prove the existence of a phase transition in a different way, and also shows that the infinite system is in a completely magnetized state at low temperatures. In Section VI we examine the spin interaction coefficients, which include all even many–body terms and are translation invariant. Our numerical results indicate that all the interaction coefficients, except the constant term, are ferromagnetic. We also discuss the question of whether the Farey and KSC interactions are the same in the limit of long chains.

## II. Definition of the Farey Spin Chain

We begin with a preliminary definition and then proceed to construct the Farey fractions.

**Definition**: *The mediant of the rational numbers* $\frac{a}{b}$ *and* $\frac{c}{d}$ *is* $\frac{a+c}{b+d}$. ◊

Note that, for $\frac{a}{b} < \frac{c}{d}$ and $\frac{a}{b}, \frac{c}{d} \in [0,1]$, $\frac{a}{b} < \frac{a+c}{b+d} < \frac{c}{d}$.

**Definition**: *The Farey fractions are defined by starting with the set* $F_1 = \left\{\frac{0}{1}, \frac{1}{1}\right\}$ *and recursively inserting mediants of each neighboring pair of fractions.* $F_n$ *denotes the sequence of fractions generated by this procedure up to and including the* $n^{th}$ *step; so that* $F_1 = \left\{\frac{0}{1}, \frac{1}{1}\right\}$, $F_2 = \left\{\frac{0}{1}, \frac{1}{2}, \frac{1}{1}\right\}$, $F_3 = \left\{\frac{0}{1}, \frac{1}{3}, \frac{1}{2}, \frac{2}{3}, \frac{1}{1}\right\}$ *etc.* ◊

$F_n$ includes all rational fractions in [0,1] with denominators d ≦ n, and others with larger denominators. Obviously $F_1 \subset F_2 \subset F_3 \subset \ldots \subset F_n \subset F_{n+1} \subset \ldots$ . We let $F_\infty$ denote the set of all fractions generated in this manner. It follows that $F_\infty = Q \cap [0,1]$.

Throughout this paper we employ a matrix representation of the elements of $F_n$. Suppose $F_m = \left\{x_0^m, x_1^m, \ldots, x_{2^{m-1}}^m\right\}$, where $x_0^m = \frac{0}{1}$, $x_{2^{m-1}}^m = \frac{1}{1}$ and $x_{j+1}^m$ is the immediate right neighbor of $x_j^m$ in $F_m$. For 0 ≦ j ≦ $2^{m-1}$, let $x_j^m = \frac{a}{b}$ and $x_{j+1}^m = \frac{c}{d}$, and suppose that the binary expansion of j is $j = (a_{m-1} \ldots a_1 a_0)_2$, $a_k \in \{0,1\}$ for 0 ≦ k ≦ m-1. Let

$$A = \begin{pmatrix} 1 & 0 \\ 1 & 1 \end{pmatrix}$$

$$B = \begin{pmatrix} 1 & 1 \\ 0 & 1 \end{pmatrix} = A^t.$$

This leads to

**Theorem 1**: $M^m(j) \equiv \prod_{k=0}^{m-1} A^{1-a_k} B^{a_k} \equiv A^{1-a_{m-1}} B^{a_{m-1}} A^{1-a_{m-2}} B^{a_{m-2}} \ldots A^{1-a_0} B^{a_0} = \begin{pmatrix} c & a \\ d & b \end{pmatrix}$, *for m ≧ 1.*

*Proof*: Theorem 1 follows from recursion relations for the Farey fractions very similar to Equation (5) of ([Ka], p. 79). ◊

*Remark*: We set $a_{m-1} = 0$, so that this product always starts with A, the only exception being when j = $2^{m-1}$ = $(100\ldots0)_2$. Then it is $BA^{m-1} = \begin{pmatrix} m & 1 \\ m-1 & 1 \end{pmatrix}$. In this case only, one employs the immediate right neighbor of $\frac{1}{1}$, $\frac{m}{m-1} > 1$. ◊

In what follows, we omit the j = $2^{m-1}$ case. Thus, the fraction $\frac{1}{1}$ is not included and the denominator 1 only appears once in $F_n$. Furthermore, $M^m(j) = AM^{m-1}(l)$, with 0 ≦ l ≦ $2^{m-2}$, $l = (a_{m-2} \ldots a_1 a_0)_2$. Therefore, for 0 ≦ j ≦ $2^{m-1}$ - 1, with $T^m(j)$ defined as $T^m(j) \equiv Tr(M^m(j))$, we have $T^m(j) = b + c = Den(x_j^m) + Num(x_{j+1}^m)$.

We are now in a position to define the partition function of the spin chain. First we extend the definition of $M^m(j)$ to include all possible products of m factors A or B, so there are now $2^m$ matrix products. Now $A = I + \sigma_-$, and similarly $B = I + \sigma_+$, where I is the (2x2) unit matrix and $\sigma_+$, $\sigma_-$ are Pauli matrices. These satisfy $\sigma_+^2 = 0 = \sigma_-^2$, $\sigma_+\sigma_- + \sigma_-\sigma_+ = I$. It follows that $M^m(j)$ is a linear combination of I, $\sigma_+$, $\sigma_-$, $\sigma_+\sigma_-$ and $\sigma_-\sigma_+$. If one exchanges A and B in the product, $\sigma_+$ and $\sigma_-$ are exchanged, but the trace remains invariant. Thus each new $T^m(j)$ is exactly the same as one from the original set.

We then interpret each $T^m(j)$ as specifying the energy of a given spin state j of a periodic chain of length m via $E_m(j) = \log T^m(j)$. The kth spin may be regarded as down or up (or, equivalently, the kth site as empty or full) according to whether $a_k = 0$ or 1, respectively.

**Definition**: *The partition function is given by*

$$Z_m(\beta) = \sum_j T^m(j)^{-\beta},$$

*where the sum extends over all $2^m$ matrix products, i.e. the first factor in* $M^m(j)$ *may be either A or B.* ◊

It follows that $Z_1 = 2^{1-\beta}$, $Z_2 = 2^{1-\beta} + 2\, 3^{-\beta}$, $Z_3 = 2^{1-\beta} + 6\, 2^{-2\beta}$, $Z_4 = 2^{1-\beta} + 4\, 6^{-\beta} + 8\, 5^{-\beta} + 2\, 7^{-\beta}$, ...

*Remark*: The A–B properties of $T^m(j)$ discussed above imply that the Farey chain exhibits up–down (equivalently, particle–hole) symmetry. It also follows immediately from the properties of the trace that the energy is invariant under cyclic translation of the spin matrices. Therefore the spin chain has translation invariant spin interactions (see Section VI). ◊

*Remark*: The partition function may be evaluated exactly for certain values of β. When β = 0, $Z_m = 2^m$, the number of states. For β = -1, $Z_m$ is the sum of the trace of all possible matrix products. Since the two operations commute, $Z_m(-1) = \mathrm{Tr}(A + B)^m$ [P]. Now A+B has eigenvalues 1 and 3, so $Z_m(-1) = 1^m + 3^m = 1 + 3^m$. One can also calculate correlation functions of the A and B matrices for these β values in a straightforward way. These simplifications may also be applied to the KSC. The β = -1 method gives an easy way to derive the results of [G-K]. ◊

*Remark*: The KSC may also be expressed using the A and B matrices. To do this, one must replace $T^m(j)$ in the formula for the partition function with $D^m(j) = b = \mathrm{Den}\left(x_j^m\right) = \left(M^m(j)\right)_{(2,2)}$ and sum only over the restricted set of matrices beginning with A. We denote the resulting partition function $Z^{K}{}_{m-1}(\beta)$. ◊

In the language of [Ka], this is the canonical partition function for a chain of length m-1 (the leading matrix A is not counted). Since all matrix elements of $M^{m}(j)$ are positive, $0 < D^{m}(j) < T^{m}(j)$. Thus the energy of each (restricted) state of the Farey chain is bounded below by the energy of the KSC. Since $Z_m$ may also be computed using the restricted sum, for $\beta > 0$ one has $Z_m(\beta) < 2Z^{K}{}_{m-1}(\beta)$. For $\beta < 0$, the inequality is reversed.

**III. Existence of the Free Energy and Phase Transition**

**Definition**: *The Farey free energy (per spin) F(β) is defined as*

$$\beta F(\beta) = \lim_{m\to\infty} \frac{-\ln[Z_m(\beta)]}{m},$$

*and the KSC free energy $F_K(\beta)$ is defined similarly, with Z replaced by $Z^K$.* ◊

**Theorem 2**: *The Farey free energy satisfies $F(\beta) = F_K(\beta)$. Thus $\beta F(\beta)$ exists for all $\beta \geqq 0$, for $\beta$ any negative integer, and exhibits a unique phase transition at $\beta = 2$.*

*Proof*: For $\beta = 0$, $\beta F(\beta) = \beta F_K(\beta) = -\ln 2$ follows immediately for either spin chain by the remark above. For the other β values, we make use of Theorem 3 below, which is independent of the present argument. With $\frac{c}{d}$ the immediate right neighbor of $\frac{a}{b}$, one sees that $c = \bar{b} + at$; $t = 0,1,2,3\ldots$, where the index t classifies the appearances of $\frac{c}{d}$ with increasing chain length m and satisfies $t < m$. Here $b\bar{b} \equiv 1 (\bmod\, a)$ and $1 \le \bar{b} \le a-1$ so that $\bar{b} < a < b$. Hence $c < (m+1)\,b$, which we make use of as follows. First note that $D^{m}(j) = b$ and $T^{m}(j) = b + c$ satisfy $D^{m}(j) < T^{m}(j) < (m+2)D^{m}(j)$. For $\beta > 0$, it follows that $Z^{K}_{m-1} > Z_m > \frac{1}{(m+2)^{\beta}} Z^{K}_{m-1}$. Taking logarithms and dividing by $-\beta m$ gives

$$\left(1-\frac{1}{m}\right)F^{K}_{m-1} < F_m < \frac{\ln(m+2)}{m} + \left(1-\frac{1}{m}\right)F^{K}_{m-1}.$$

Our results for positive β are then established by taking the limit $m \to \infty$, and using the rigorous results that $\beta F_K(\beta)$ exists for all $\beta > 0$ [Kd] and exhibits a unique phase transition at $\beta = 2$ [C-K]. For $\beta < 0$ the same inequality holds for $F_m$; thus since $\beta F_K(\beta)$ is also known to exist for β any negative integer [C-K], the rest of the theorem is proved. ◊

*Remark*: Another way of establishing the phase transition and gaining information on the behavior of the system at large β is given in Theorem 4. ◊

*Remark*: Note that F(-1) = ln3 by the calculation above. Making use of [C-K] gives the same result for $F_K(-1)$. ◊

Figures 1 and 2 illustrate the free energy and energy fluctuation per spin ΔE (which is proportional to the specific heat) obtained by exact enumeration for chains up to length m = 16. It is clear that the convergence of the free energy with length is much slower at large β, where the system is known to be completely magnetized in the infinite m limit (see Section V), than for small β values. In addition, Figure 1a shows that the approach to the limit is non–monotonic for small chain lengths, for β = 1, which is not the case at larger β values (cf. Figure 1b). Finally, Figure 2 shows a peak in ΔE near β = 2, consistent with the phase transition there. ΔE increases much more rapidly with length at the peak (β = 2) than at nearby β values.

**IV. Number of States**

In this Section we consider the summed number of states of the spin chain, that is, the number of states of a given energy regardless of chain length. To begin, we derive an expression for the number of immediate right neighbors of a given Farey fraction.

**Definition**: *If m is the smallest positive integer such that* $\frac{a}{b} \in F_m$, *then we call m the conductor of* $\frac{a}{b}$. *We write* $m = \mathrm{cond}\left(\frac{a}{b}\right)$. ◊

Let $\mathrm{cond}\left(\frac{a}{b}\right) = m$, and let $\frac{x'}{y'} < \frac{a}{b} < \frac{x}{y}$ so that $\frac{x'}{y'}, \frac{x}{y} \in F_{m-1}$ are the immediate neighbors of $\frac{a}{b}$ in $F_m$. Then we have

$$a = x + x' \quad ; \quad 0 \le x, x' < a$$
$$b = y + y' \quad ; \quad 0 \le y, y' < b$$

and

$$bx - ay = (y + y')x - (x + x')y = y'x - x'y.$$

Similarly,

$$bx' - ay' = (y + y')x' - (x + x')y' = yx' - xy'.$$

We conclude, by induction on m, that

$$bx - ay = 1$$

and $$bx' - ay' = -1,$$

so that

$$bx \equiv \ 1(\bmod a)\,;\, 0 \le x < a \qquad (1)$$

and $$ay \equiv -1(\bmod b)\,;\, 0 \le y < b. \qquad (2)$$

Suppose $\bar{b}$ is the unique multiplicative inverse of b (mod a), i.e. the unique solution of (1). Then

$$b\bar{b} - ay = 1 \text{ so that}$$

$$y = \frac{b\bar{b} - 1}{a}.$$

**Definition**: *Let* $N_R\left(\frac{a}{b}\right)$ *denote the set of all immediate right neighbors of* $\frac{a}{b}$ *in* $F_m$*, for* $m \geqq$ *cond(a/b).* ◊

The above provides a proof of

**Theorem 3**: *.The immediate right neighbor of* $\frac{a}{b}$ *at step* $m = \text{cond}\left(\frac{a}{b}\right)$ *is* $\frac{\bar{b}}{\left(\frac{b\bar{b}-1}{a}\right)}$ *and*

$$N_R\left(\frac{a}{b}\right) = \left\{ \frac{\bar{b} + at}{\left(\frac{b\bar{b} + abt - 1}{a}\right)} \middle| t = 0,1,2,\ldots \right\}..$$

*Remark*: If $m = \text{cond}\left(\frac{a}{b}\right)$, then in $F_m$ , one can show that the immediate left neighbor of $\frac{a}{b}$ is $\frac{a - \bar{b}}{\bar{a}}$ and its immediate right neighbor is $\frac{\bar{b}}{b - \bar{a}}$, where $\bar{a}$ is the unique multiplicative inverse of a (mod b). The form for the right neighbor is a consequence of the fact that $a\bar{a} + b\bar{b} = 1 + ab$. ◊

The next step is to find the number of states with a given energy, regardless of chain length. We proceed to count the number of solutions to

$$\mathrm{Tr}(X) = n, \tag{3}$$

where n > 2 and X is any finite product of A's and B's beginning with A as described above.

We count the contribution of $\frac{a}{b}$ to the left side of (3) by keeping track of the immediate right neighbors of $\frac{a}{b}$. For this, we consider Theorem 3 and

$$b + \bar{b} + at = n\ ; t = 0,1,2,... \tag{4}$$

which is equivalent to

$$b^2 + b\bar{b} + abt = bn$$

or $$b^2 - bn + 1 \equiv 0 \pmod{a}.$$

Therefore we look for the divisors a of $bn - b^2 - 1$ for $1 \le a < b$. There are $d_b\left(bn - b^2 - 1\right)$ of these, where $d_b(m)$ is the number of positive divisors of m that are less than b. It immediately follows that

**Theorem 4**: *The number of solutions of (3) is* $\Phi(n) = \sum_{b=1}^{n-1} d_b\left(bn - b^2 - 1\right)$.

Similar quantities are considered in [H-T].

*Remark*: We can put an upper bound on $\Phi(n)$ by considering (4). If we ignore the restrictions, the number of solutions to it is just the number of unordered partitions of n into three positive integers. Therefore

$$\Phi(n) \le \binom{n+2}{2} = \frac{1}{2}(n+2)(n+1) \sim \frac{1}{2}n^2,$$

where the asymptotic form applies as $n \to \infty$. ◊

**Conjecture**: *As* $n \to \infty$, $\Phi(n) \sim \frac{1}{2} n \log n$. ◊

Consider the fact that $\sum_{m=1}^{n} d(m) \sim n \log n$, which implies that for large n the average value of d is log n, and the DDT theorem $\sum_{m=1}^{n} \frac{d_{m^u}(m)}{d(m)} \sim \frac{2}{\pi}\sin^{-1}(\sqrt{u}) \cdot n$ [D-D-T]. Now replace the quantities d(m) and $d_b(m)$ by the averaged functions d(x) and d(x;y) where the latter is the averaged number of divisors of $y \leq x$. Taken together, the quoted results suggest that $\int_1^n d(x^u;x)dx \sim \frac{2}{\pi}\sin^{-1}(\sqrt{u}) \cdot n\log n$, and more generally the averaged quantity satisfies

$$d(g;x) \sim \frac{2}{\pi}\sin^{-1}(\sqrt{\frac{\log g}{\log x}}) \cdot \log n .$$

for large $x \leqq n$. The argument $\frac{\log g}{\log x}$ is consistent with the fact that divisors are uniformly distributed on a log scale ([H-T], p. 62). The asymptotic form of the number of states then follows

$$\Phi(n) \sim \int_1^{n-1} d(b;bn - b^2 - 1)db$$

$$\sim \frac{2}{\pi}\int_1^{n-1} \sin^{-1}(\sqrt{\frac{\log b}{\log b(n-b)}})db \cdot \log n$$

Evaluating the integral for large n, we find $\sin^{-1}(\frac{1}{\sqrt{2}})n = \frac{\pi}{4}n$, which leads immediately to the conjectured formula.

**V. Thermodynamic Consequences**

The results for the (summed) number of states in Section IV may be used to derive some consequences for the thermodynamics. We first express the partition function as

$$Z_m(\beta) = 2^{1-\beta} + 2\sum_{n=3}^{\infty}\frac{\Phi_m(n)}{n^\beta},$$

where $\Phi_m(n)$ is the number of solutions of (3) for fixed chain length m, i.e. the number of states. The factor 2 appears in front of the summation of the preceding equation since (3) refers to the restricted set of matrices beginning with A. The summed number of states is then $\Phi(n) = \sum_{m=1}^{\infty}\Phi_m(n)$. If we define $Z'_m(\beta) \equiv Z_m(\beta) - 2^{1-\beta}$ and use the conjectured asymptotic form of $\Phi(n)$, then

$$\sum_{m=1}^{\infty} Z'_m(\beta) = 2\sum_{n=3}^{\infty} \frac{\Phi(n)}{n^{\beta}}$$

$$= \sum_{n=1}^{\infty} \frac{n\log n}{n^{\beta}} + 2\sum_{n=1}^{\infty} \frac{\varepsilon(n)}{n^{\beta}} - \frac{2^{2-\beta}}{3}\log 2$$

$$= -\zeta'(\beta-1) + 2\tilde{\varepsilon}(\beta) - \frac{2^{2-\beta}}{3}\log 2,$$

where $\varepsilon(n)$ corrects the asymptotic form, $\tilde{\varepsilon}(\beta)$ is its Dirichlet transform, and $\zeta'$ is the derivative of the Riemann $\zeta$–function. If we assume that $\tilde{\varepsilon}(\beta)$ is regular, then a singularity in the thermodynamics for $\beta = 2$ follows from the pole in the Riemann function at $\beta-1 = 1$. We have not been able to prove this assumption, however.

**Theorem 4**: *$Z'_m(\beta) \to 0$ as $m \to \infty$, so that $Z_m(\beta) \to 2^{1-\beta}$, and the free energy $F(\beta) = 0$ for $\beta > 3$.*
*Proof*: We use the upper bound on $\Phi(n)$ derived above. Now

$$\sum_{m=1}^{\infty} Z'_m(\beta) = \sum_{n=3}^{\infty} \frac{\Phi(n)}{n^{\beta}} \le \sum_{n=3}^{\infty} \frac{(n+2)(n+1)}{n^{\beta}}.$$

The summation in this equation is finite for $\beta > 3$. ◊

*Remark*: Since $\beta F(\beta)$ is finite at $\beta = 0$, Theorem 4 establishes the existence of a phase transition (singularity in $\beta F(\beta)$) in a different way than used in Section III. In addition, since the partition function for long chains is given by the sum over the two lowest energy ($j = 0$ or $2^m-1$) states only, it shows that the limiting chain is in a completely magnetized state (all spins up or all spins down) at low temperatures. Similar behavior holds, by construction, for the KSC, with the partition function approaching a ratio of Riemann $\zeta$–functions for $\beta > 2$ as $m \to \infty$. For the Farey chain, replacing the upper bound by our conjectured asymptotic form for $\Phi(n)$, leads to $F(\beta) = 0$ and $Z_m(\beta) \to 2^{1-\beta}$ for $\beta > 2$. ◊

## VI. Spin Interaction Coefficients

We define the spin interaction coefficients $J_m(t)$ as follows [Kb]. Let t, $0 \leqq t \leqq 2^{m-1}$, specify the coefficient, and suppose that the binary expansion of t is $t = (b_{m-1}\ldots b_1 b_0)_2$, $b_k \in \{0,1\}$. Let $j \cdot t = \sum_{i=0}^{m-1} a_i b_i$. Then

$$J_m(t) \equiv -\frac{1}{2^m}\sum_j (-1)^{j\cdot t} E_m(j),$$

where $E_m(j)$ is the energy of configuration j for a chain of length m, defined above. In this notation, the energy of any configuration is given via a sum over spin clusters

$$E_m(j) = -\sum_t (-1)^{j \cdot t} J_m(t).$$

Note that each factor $s_i \equiv (-1)^{a_i} \in \{-1,+1\}$ in each term may be interpreted as a spin at site i on the chain. The spin $s_i$ is present or not in a given term according to whether $b_i = 1$ or $b_i = 0$, respectively. More explicitly,

$$E_m(j) = -\sum_{\{b_i\}} \left[ \prod_{i=0}^{m-1} (s_i)^{b_i} \right] J(\{b_i\}).$$

Each term thus defines a spin cluster, i.e. a set of sites for which $b_i = 1$. The sites in a given cluster may be adjacent to one another or separated by $b_i = 0$ sites with no spins present.

**Lemma**: *The interaction coefficient $J_m(0)$ satisfies*

$J_m(0) = -\frac{1}{2^m} \sum_j \ln T_m(j) \geq -m \ln(\frac{\sqrt{5}+1}{2}) = -(0.48121...)m$ *for large m.*

*Proof*: By considering the generation of the Farey fractions $F_m$, it is clear that each numerator and denominator is bounded above by the Fibonacci number $F_{m+1}$. Therefore $T^m(j) \leq 2F_{m+1} \sim \frac{2}{\sqrt{5}} \left( \frac{1+\sqrt{5}}{2} \right)^{m+1}$, as m -> ∞. ◊

We have calculated $J_m(0)$ by computer for chains up to m = 16 as illustrated in Figure 3. The results indicate that it approaches -(a m + b), with a = 0.3962, for both the Farey spin chain and KSC.

Since $J_m(0)$ is also (minus) the average energy for a chain of length m, we considered the fluctuation of the energy, i.e.

$$\sigma_m{}^2 = \frac{1}{2^m} \sum_j (\ln T^m(j))^2 - (\frac{1}{2^m} \sum_j \ln T^m(j))^2.$$

Numerically, from calculations on chains up to length 16, $\sigma_m$ appears to approach 0.019 m as m -> ∞. The corresponding quantity for the KSC apparently approaches 0.014 m. Given the numerical uncertainties, it is possible that the asymptotic value is the same for both chains.

Note that, since $T_m \geqq 2$, $J_m(0) < 0$. By the up–down symmetry discussed above, it follows that $J_m(t) = 0$ whenever the cluster defined by t contains an odd number of spins, i.e. when $\sum_i b_i = \text{odd}$, so only even interaction coefficients are non–zero. Furthermore, $J_m(t)$ exhibits cyclic symmetry in t, i.e. it is invariant under translation of the spin cluster, due to the invariance of the trace under cyclic translation of the spin matrices mentioned above.

Our computer results for short chains verify all the exact behavior mentioned in the preceding paragraph. In addition, we find several interesting features. For all non–zero clusters (with an even number of spins) $J_m(t) > 0$, so all interaction coefficients (except t = 0) are ferromagnetic. Similar behavior occurs for the KSC in what is referred to as the grand canonical ensemble [Kb]. We find that as m increases, each such $J_m(t)$ apparently approaches a finite limit. An example is shown in Figure 4. Denoting by J(t) the limit of $J_m(t)$ as $m \to \infty$, we find the values listed in Table I. Note that the (approximate) numbers for the pair interaction coefficients (t = (10...01000...) ) are consistent with a decrease of J by a factor of 1/2 for each increase of separation of the two spins in the cluster by one site. Generally, J appears to decrease with the number of spins in the cluster and the distance between the spins. In addition we find that in general $J_m(t_r) \neq J_m(t)$, where $t_r \equiv (b_0 b_1 \ldots b_{m-1})_2$, so the interactions do not exhibit the KSC reflection symmetry.

Theorem 2 establishes that the Farey spin chain and KSC have the same free energy. This raises the question as to whether J(t) is in some sense the same in both cases. This is a more complicated issue, in part because the KSC is not translation invariant. Our numerical results are consistent with the J(t) being the same, at least for KSC interactions with spin clusters far from the edges, i.e. with t of the form 0...0p0...0, with p = 1...1 fixed and each string of 0s of length proportional to m. For t of the form p0...0, so the cluster remains at one edge, the J(t) values are certainly different. Further, $J_m(t)$ for the KSC is rigorously known to exhibit the folowing behavior for $m \to \infty$: $J_m(t) \to 0$ when t has an odd number of spins, it is small unless the length of p is small, and it has translation invariance [Ka]. All of these are consistent with what we find for the Farey chain, as described above.

Unfortunately, the bounds relating $D^m(j)$ and $T^m(j)$ used in the proof of Theorem 2 are not strong enough to establish equality, since each term in $J_m(t)$ will be bounded above and below by the corresponding term in $J_m{}^K(t)$ plus (depending on sign) a possible term of magnitude $\ln(m+2)\, 2^{-m}$, and on summation the latter gives rise to a divergent contribution as $m \to \infty$. However, one can draw some conclusion about $J_m(0)$ and $\sigma_m$ in this way. We assume, consistent with the Lemma and numerical results above, that both quantities are proportional to m in this limit. It is then easy to show that the respective coefficients must be the same for either spin chain, as suggested by the numerics.

*Ackndowledgements*. We acknowledge stimulating and useful interactions with P. Contucci, A. Knauf and I. Peschel, and thank A. Knauf for a computer program.

Table I

| t | J(t) |
|---|---|
| 11000... | 0.131 |
| 101000... | 0.0612 |
| 1001000... | 0.0291 |
| 10001000... | 0.0141 |
| 100001000... | 0.0068 |
| 1000001000... | 0.0033 |
| 1111000... | 0.0081 |
| 11011000... | 0.0028 |
| 110011000... | 0.0011 |
| 11101000... | 0.004 |

**Captions:**

Table I: Interaction coefficients.

Figure 1: Free energy $F_m \equiv \frac{-\ln Z_m}{m\beta}$ vs. m for Farey spin chain at (a) $\beta = 1$ and (b) $\beta = 4$.

Figure 2: Fluctuation of the energy per spin $\Delta E \equiv \frac{1}{m}\left(\left\langle E_m(j)^2\right\rangle - \left\langle E_m(j)\right\rangle^2\right)$ $\Delta E \equiv \frac{1}{m}\left(\left\langle E_m(j)^2\right\rangle - \left\langle E_m(j)\right\rangle^2\right)$ vs. $\beta$ for the Farey spin chain with m = 16.

Figure 3: $J_{m+1}(0)$ - $J_m(0)$ vs. m for the Farey chain (stars) and KSC (diamonds).

Figure 4: Pair interaction $J_m(11000...)$ vs. m for the Farey chain.